\documentclass[runningheads,a4paper]{llncs}

\usepackage{wrapfig}
\usepackage[dvipsnames]{xcolor}
\usepackage{tabularx}
\usepackage{booktabs}
\usepackage{amssymb,amsfonts,amsmath}
\usepackage{url}
\usepackage{framed}

\definecolor{light-gray}{gray}{0.9}
\usepackage[framemethod=TikZ]{mdframed}

\mdfsetup{skipabove=5pt,skipbelow=3pt}
\usepackage{lipsum}
\mdfdefinestyle{MyFrame}{%
    linecolor=black,
    outerlinewidth=0.15pt,
    roundcorner=3pt,
    innertopmargin=2pt,
    innerbottommargin=2pt,
    innerrightmargin=4pt,
    innerleftmargin=4pt,
    backgroundcolor=light-gray}
    
\mdfdefinestyle{MyFrameSimple}{%
    linecolor=black,
    outerlinewidth=0.15pt,
    roundcorner=3pt,
    innertopmargin=2pt,
    innerbottommargin=2pt,
    innerrightmargin=4pt,
    innerleftmargin=4pt}

\newcommand{\sectopic}[1]{\vspace*{.2em}\par\noindent{\textit{\bfseries #1}}}

\begin{document}

\title{Using Language Models for Enhancing the Completeness of Natural-language Requirements}
\titlerunning{Using Language Models for Enhancing the Completeness of Requirements}

\author{Dipeeka Luitel\and Shabnam Hassani \and Mehrdad Sabetzadeh}

\authorrunning{D. Luitel, S. Hassani, M. Sabetzadeh}

\institute{University of Ottawa, Ottawa ON, K1N 6N5 Canada\\
\email{\{Dipeeka.Luitel,\,S.Hassani,\,M.Sabetzadeh\}@uottawa.ca}}

\maketitle

\begin{abstract}
\textbf{[Context and motivation]} Incompleteness in natural-\linebreak language requirements is a challenging problem.
\textbf{[Question/problem]} A common technique for detecting incompleteness in requirements is checking the requirements against external sources. With the emergence of language models such as BERT, an interesting question is whether language models are useful external sources for finding potential incompleteness in requirements. 
\textbf{[Principal ideas/results]} We mask words in requirements and have BERT's masked language model (MLM) generate contextualized predictions for filling the masked slots. 
We simulate incompleteness by withholding content from requirements and measure BERT's ability to predict terminology that is present in the withheld content but absent in the content disclosed to BERT.
\textbf{[Contribution]} BERT can be configured to generate multiple predictions per mask. Our first contribution is to determine how many predictions per mask is an optimal trade-off between effectively discovering omissions in requirements and the level of noise in the predictions. Our second contribution is devising a machine learning-based filter that post-processes predictions made by BERT to further reduce noise. We empirically evaluate our solution over 40 requirements specifications drawn from the PURE dataset~\cite{ferrari2017pure}. Our results indicate that: (1) predictions made by BERT are highly effective at pinpointing  terminology that is missing from requirements, and (2) our filter can substantially reduce noise from the predictions, thus making BERT a more compelling aid for improving completeness in requirements.

\vspace*{-.5em}
\end{abstract}
\vspace*{-2em}
\section{Introduction}\label{sec:intro}
Improving the completeness of requirements is an important yet challenging problem in requirements engineering (RE)~\cite{Zowghi:03}. 
The RE literature distinguishes two notions of completeness~\cite{Zowghi:03b}: (1)~\emph{Internal} completeness is concerned with requirements being closed with respect to the functions and qualities that one can infer exclusively from the requirements. (2)~\emph{External} completeness is concerned with ensuring that requirements are encompassing of all the information that external sources of knowledge suggest the requirements should cover. These external sources can be either people (stakeholders) or artifacts, e.g., higher-level requirements and existing system descriptions~\cite{arora2019empirical}. External completeness is a relative measure, since the external sources may be incomplete themselves or not all the relevant external sources may be known~\cite{Zowghi:03b}. Although external completeness cannot be defined in absolute terms, relevant external sources, when available, can be useful for detecting missing requirements-related information.

When requirements and external sources of knowledge are textual, one can leverage natural language processing (NLP) for computer-assisted checking of external completeness. For example, Ferrari et al.~\cite{ferrari2014measuring} use NLP to check completeness against stakeholder-interview transcripts. And, Dalpiaz et al.~\cite{dalpiaz2018pinpointing} use NLP alongside visualization to identify differences among stakeholders' viewpoints; these differences are then investigated as potential incompleteness issues.

With (pre-trained) language models, a new opportunity arises for NLP-based improvement of external completeness in requirements: Using self-supervised learning, language models  have been trained on very large corpora of textual data, e.g., millions of Wikipedia articles. This raises the prospect that \emph{a language model  can serve as an external source of knowledge for completeness checking}. In this paper, we explore a specific instantiation of this idea using BERT~\cite{devlin2018bert}.

\begin{figure}[!t]
\centering
    \includegraphics[width=1\linewidth]{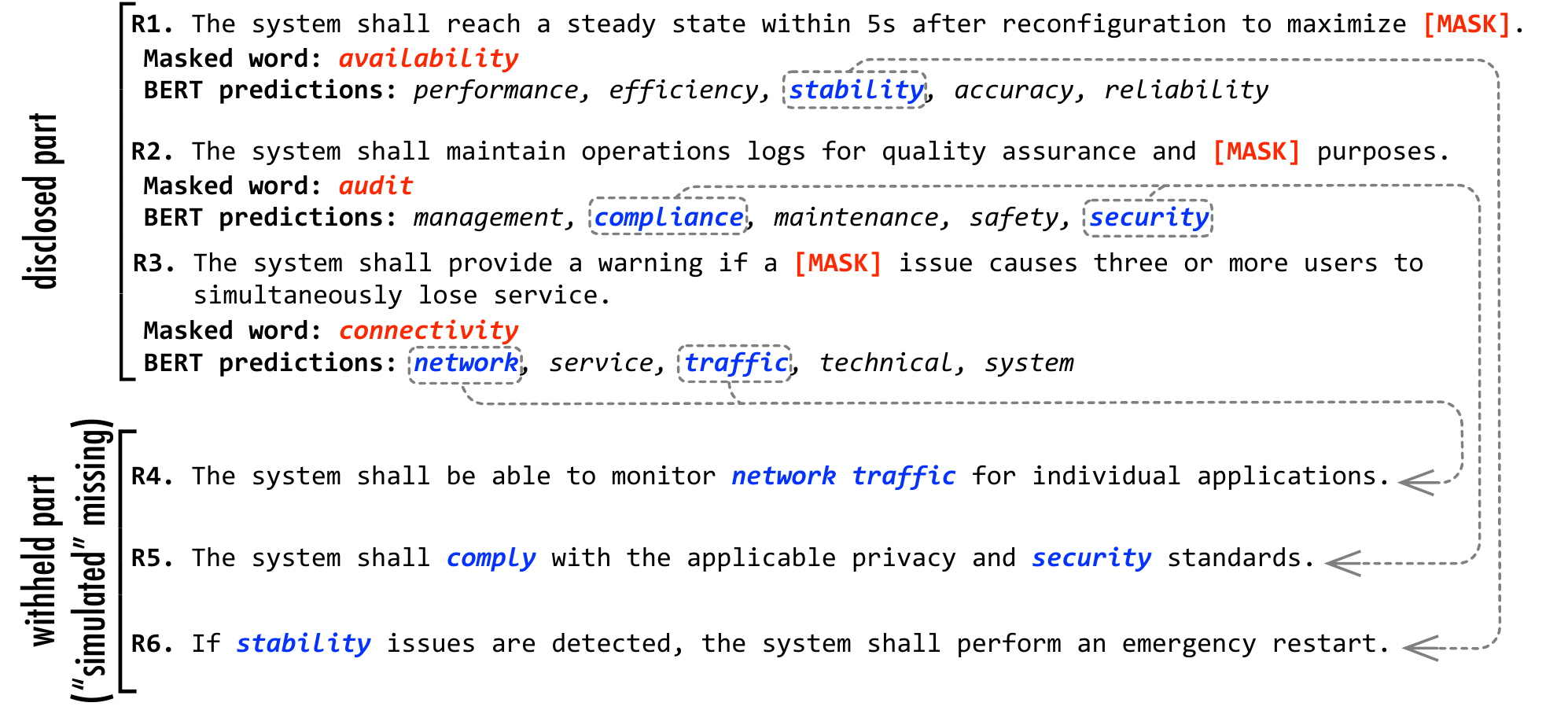}
\vspace*{-2.2em}
\caption{Illustrative requirements specification split into a \emph{disclosed} and a \emph{withheld} part. The withheld part \emph{simulates} requirements omissions. Masking words in the disclosed part and having BERT make predictions for the masks reveals some terms that appear only in the withheld part.}
\vspace*{-1.6em}
\label{fig:example}
\end{figure}

BERT has been trained to predict masked tokens by finding words or phrases that most closely match the surrounding context. To illustrate how BERT can help with completeness checking of requirements, consider the example in Fig.~\ref{fig:example}. In this example, we have masked one word, denoted \texttt{[MASK]}, in each of requirements $\textsf{R1}$, $\textsf{R2}$ and $\textsf{R3}$. We have then had BERT make five predictions for filling each masked slot. For instance, in $\textsf{R1}$, the masked word is \emph{availability}. The predictions made by BERT are: \emph{performance}, \emph{efficiency}, \emph{stability}, \emph{accuracy}, and \emph{reliability}. As seen from the figure, one of these predictions, namely \emph{stability}, is a word that appears in $\textsf{R6}$. Similarly, the predictions that BERT makes for the masked words in $\textsf{R2}$ and $\textsf{R3}$ (\emph{audit} and \emph{connectivity}) reveal new terminology that is present in $\textsf{R4}$ and $\textsf{R5}$ (\emph{network}, \emph{traffic}, \emph{comply} and \emph{security}).

In the above example, \textbf{\emph{if requirements $\textsf{R4}$--$\textsf{R6}$ were to be missing, BERT's predictions over $\textsf{R1}$--$\textsf{R3}$ would  provide useful cues about some of the missing terminology.}}

\sectopic{Contributions.} We need a strategy to study how well BERT predicts relevant terminology that is absent from requirements. To this end, we \emph{simulate} missing information by randomly withholding a portion of a given requirements specification. We disclose the remainder of the specification to BERT for obtaining masked-word predictions. In our example of Fig.~\ref{fig:example}, the disclosed part would be requirements $\textsf{R1}$--$\textsf{R3}$ and the withheld part would be requirements $\textsf{R4}$--$\textsf{R6}$. BERT can be configured to generate multiple predictions per mask. Our \textbf{\emph{first contribution}} is to determine how many predictions per mask is an optimal trade-off between effectively discovering simulated omissions and the amount of unuseful predictions (noise) that BERT generates. 

We observe that a large amount of noise would result if predictions by BERT are to achieve good coverage of requirements omissions. Some of the noise is trivial to filter. For instance, in the example of Fig.~\ref{fig:example}, one can dismiss the predictions of \emph{service} and \emph{system} (made over \textsf{R3}); these words already appear in the disclosed portion, thereby providing no  cues about missing terminology. Furthermore, one can dismiss words that carry little meaning, e.g., ``any'', ``other'' and ``each'', should such words appear among the predictions. After applying these obvious filters, one would still be left with considerable  noise. Our \textbf{\emph{second contribution}} is a machine learning-based filter that post-processes predictions by BERT to strike a better balance \hbox{between noise and useful predictions.}

Our solution development and evaluation is based on 40 requirements specifications from the PURE dataset~\cite{ferrari2017pure}. These specifications contain over 23,000 sentences combined.
To facilitate replication and further research, we make our implementation and evaluation artifacts publicly available~\cite{Luitel:22}.

\vspace*{-.5em}
\section{Background}\label{sec:background}
\vspace*{-.2em}
Below, we review the background for our work, covering the NLP pipeline, language models, word embeddings, machine learning and corpus extraction.

\sectopic{NLP Pipeline.} \label{subsec:NLP Pipeline}
    Natural language processing (NLP) is usually performed using a pipeline of modules~\cite{jurafsky2009speech}. In this paper, we apply an NLP pipeline  composed of tokenizer, sentence splitter, part-of-speech (POS) tagger and lemmatizer modules. The tokenizer demarcates the tokens of the text. The sentence splitter divides the text into sentences. The POS tagger assigns a POS tag to each token in each sentence. The lemmatizer maps each word in the text to its lemma form. For example, the lemma for both ``running'' and ``ran'' is ``run''.
    We use the annotations produced by the NLP pipeline for several purposes, including the identification and lemmatization of terms in requirements documents as well as processing predictions made by BERT in \hbox{their surrounding context.}
     
\sectopic{Language Models.}\label{subsec:LM}
    Recent NLP approaches heavily rely on deep learning, and in particular, transfer learning~\cite{devlin2018bert}. Bidirectional Encoder Representations from Transformers (BERT) is a pre-trained language model using two unsupervised tasks: Masked Language Model (MLM) and Next Sentence Prediction (NSP). BERT Base and BERT Large are two types of the BERT model. BERT Large, while generally more accurate, requires more computational resources. To mitigate computation costs, we employ BERT Base. BERT Base has 12 encoder layers with a hidden size of 768 and $\approx$110 million trainable parameters.
    BERT models take capitalization of words into consideration and can be either cased or uncased. For BERT uncased, the text has been lower-cased before tokenization, whereas in BERT cased, the tokenized text is the same as the input text. Previous RE research suggests that the cased model is preferred for analyzing requirements~\cite{hey2020norbert,ezzini2022automated}. We thus use the cased model in this paper.
    
\sectopic{Word Embeddings.}\label{subsec:WE}
    In our work, we need a semantic notion of similarity that goes beyond lexical equivalence and allows us to further identify closely related terms; examples would be (i)~``key'' and ``unlock'', and (ii)~``encryption'' and ~``security''. For this, we use cosine similarity over \emph{word embeddings}.
    Word embeddings are mathematical representations of words as dense numerical vectors capturing syntactic and semantic regularities~\cite{mikolov2013linguistic}. We employ GloVe's pre-trained model~\cite{pennington2014glove}. This choice is motivated by striking a trade-off between accuracy and efficiency. BERT also generates word embeddings; however, these embeddings are expensive to compute and do not scale well when a large number of pairwise term comparisons is required.
     
\sectopic{Machine Learning (ML).}\label{subsec:ML}
    We use supervised learning, more specifically classification, to identify the relevant predictions made by BERT. Our features for learning and our process for creating labelled data are discussed in Sections~\ref{sec:approach}~and~\ref{sec:eval}, respectively. Classification models have a tendency to predict the more prevalent class(es)~\cite{WittenFrankEtAl17}. In our context, non-relevant terms outnumber relevant ones. We under-sample the majority class (i.e., non-relevant) to counter imbalance in our training set and thereby reduce the risk of filtering useful information~\cite{berry2017}. To further reduce this risk, we additionally employ cost-sensitive learning (CSL)~\cite{WittenFrankEtAl17}. CSL enables us to assign a higher penalty to relevant terms being filtered than non-relevant terms \hbox{being classified as relevant.}
     
\sectopic{Domain-corpus Extraction.}\label{subsec:DCE}
    Domain-specific corpora are useful resources for improving the accuracy of automation in RE~\cite{ezzini2021using}. When such corpora do not exist a priori, they can be constructed using domain documents from sources such as Wikipedia, books and magazines~\cite{ezzini2021using,cui2008corpus,ferrari2017detecting,ezzini2022wikidominer}. In our work, we require statistical information from a domain-specific corpus to better determine  relevance of terms predicted by BERT. For this purpose, we employ the WikiDoMiner corpus extractor~\cite{ezzini2022wikidominer}.
    WikiDoMiner is a fully automated tool that gathers domain knowledge for an input requirements specification by crawling Wikipedia. The tool extracts keywords from the input specification and assembles a set of Wikipedia articles relevant to the terminology and thus the \hbox{domain of the specification.}
    
\vspace*{-.5em}
\section{Approach}\label{sec:approach}
\begin{figure}[!t]
\centering
    \includegraphics[width=1\linewidth]{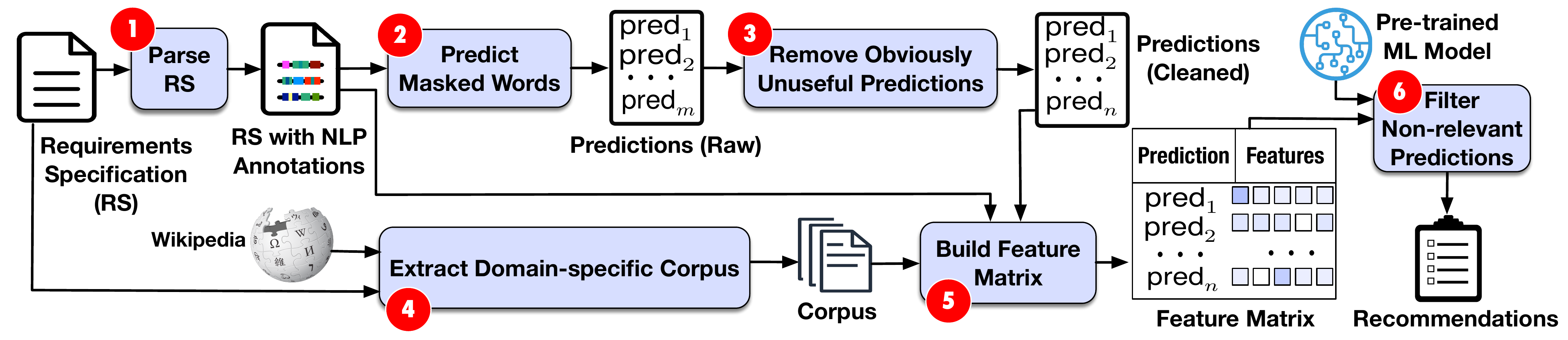}
\vspace*{-1.8em}
\caption{Approach Overview.}\label{fig:approach}
\vspace*{-1.5em}
\end{figure}

Fig.~\ref{fig:approach} provides an overview of our approach. The input to the approach is a (textual) requirement specification (RS). The approach has six steps. The first step is to parse the RS. The second step is to generate predictions for masked words using BERT. The third step is to remove the predictions that  provide little or no additional information.
The fourth step is to construct a domain-specific corpus for the given RS. Using this corpus and
the results from Step~1, the fifth step is to build a feature matrix for ML-based filtering of non-relevant terms from predictions by BERT. The sixth and last step is to feed the computed
feature matrix to a (pre-trained) classifier in an attempt to remove noise (non-relevant
words) from the predictions. The output of the approach is a list of
recommended terms that are likely relevant to the RS but are \hbox{currently absent from it.}

\sectopic{Step~1)~Parsing RS using NLP.}
The RS is fed to an NLP pipeline. The pipeline first annotates the sentences in the RS using a sentence splitter. A sentence annotation does not necessarily demarcate a grammatical sentence but rather what the sentence splitter finds to be a sentence. Next, each word in each sentence is annotated with a POS tag and \hbox{the lemma form of the word.}

\sectopic{Step~2)~Obtaining Predictions from BERT.}
Our approach loops through each sentence of the annotated RS obtained from Step 1. It masks, one at a time, each word that has a POS tag of ``noun'' or ``verb''. The resulting sentence with a single masked word is fed to BERT in order to obtain a configurable number of predictions for the masked word. We focus on nouns and verbs because noun phrases and verb phrases are the main meaning-bearing elements of sentences~\cite{arora2019empirical}. In our illustration of Fig.~\ref{fig:example}, we had BERT generate five predictions per masked word. As we argue empirically in our evaluation of Section~\ref{sec:eval}, for our purposes, our recommendation is 15 predictions per masked word. For each prediction, BERT provides a probability score indicating its confidence in the prediction. We retain the probability scores for use in Step 5 of the approach.

\sectopic{Step~3)~Removing Obviously Unuseful Predictions.}
We discard predictions that offer little or no additional information. Specifically, we remove predictions whose lemma is already present in the RS; such predictions provide no new hints about potentially missing terminology. We further remove predictions that  either (1) are among the top 250 most common words in English, or (2) belong to the union of Berry et al.'s~\cite{Berry:03} and Arora et al.'s~\cite{arora2015tse}~\cite{arora2017tse} sets of vague words and stopwords in requirements. The output of this step is a list of predictions cleared of obviously unuseful terms.

\sectopic{Step~4)~Generating Domain-specific Corpus.}
Using WikiDoMiner~\cite{ezzini2022wikidominer} (introduced in Section~\ref{sec:background}), we automatically extract from Wikipedia a domain-specific corpus for the input RS. WikiDoMiner has a depth parameter that controls the expansion of the corpus. When this parameter is set to zero, we obtain a corpus of articles containing a direct match to the key phrases in the RS. Increasing the depth generates larger corpora, with each level further expanding the sub-categories of Wikipedia articles included. In our work, we restrict our search to direct article matches (i.e., \hbox{\emph{depth = 0}}). This enables quick corpus generation and further scopes terminology expansion to what is most immediately pertinent to the RS at hand. In our exploratory investigation, we observed that larger depth values significantly increase corpus size, diluting its domain-specificity and in turn reducing the effectiveness of filtering \hbox{in Step~6 of the approach.}

\sectopic{Step~5)~Building Feature Matrix for Filtering.}
For each prediction from Step 3, we compute a feature vector as input for a ML-based classifier that decides whether the prediction is ``relevant'' or ``non-relevant'' to the input RS. Our features are listed and explained in Table~\ref{tbl:featurestable}. The main principle behind our feature design has been to keep the features generic and in a normalized form. Being generic is important because we do not want the features to rely on any particular domain or terminology. Having the features in a normalized form is important for allowing labelled data from multiple documents to be combined for training, and for the resulting ML models to be applicable to unseen documents. The output of this step is a feature matrix where each row represents a prediction (from Step 3) and each column represents a feature as defined in Table~\ref{tbl:featurestable}.

\begin{table}[!t]
\caption{Features for Learning Relevance and Non-relevance of Predictions by BERT.}
\label{tbl:featurestable}
\fontsize{9}{9.2}\selectfont
\begin{tabularx}{\columnwidth}{p{3em} X}
  \toprule
  \textbf{ID} & 
  \textbf{Type (T), 
  Definition (D) and
  Intuition (I)}\\
  \midrule
  \hline
  \vphantom{\large M}F1 &
  \textbf{(T)} Nominal \textbf{(D)} POS tag of the masked word (noun or verb).  \textbf{(I)} This feature is helpful if nouns and verbs happen to influence relevance in different ways.\\

  F2 &
  \textbf{(T)} Nominal \textbf{(D)} POS tag of the prediction; this is obtained by replacing the masked word with the predicted word and running the NLP pipeline on the resulting sentence. \textbf{(I)} The intuition is similar to F1, except that predictions are not necessarily nouns or verbs and can, e.g., be adjectives or adverbs.\\
  
  F3 &
  \textbf{(T)} Nominal (Boolean) \textbf{(D)} True if F1 and F2 match; otherwise, False. \textbf{(I)} A mismatch between F1 and F2 could be an indication that the prediction is non-relevant. \\ 
 
  F4 &
  \textbf{(T)} Numeric \textbf{(D)} Length (in characters) of the masked word. \textbf{(I)} Words that are too short may give little information. As such, predictions resulting from masking short words could be non-relevant. \\ 

  F5 &
  \textbf{(T)} Numeric \textbf{(D)} Length (in characters) of the prediction. \textbf{(I)} Predictions that are too short could be non-relevant. \\ 

  F6 &
  \textbf{(T)} Numeric \textbf{(D)} $\min\text{(F4, F5)}/\max\text{(F4, F5)}$. \textbf{(I)} A small ratio (i.e., a large difference in length between the prediction and the masked word) could indicate non-relevance. \\
  
  F7 &
  \textbf{(T)} Numeric \textbf{(D)} The confidence score that BERT provides alongside the prediction. \textbf{(I)} A prediction with a high confidence score could have an increased likelihood of being relevant. \\ 

  F8 &
  \textbf{(T)} Numeric \textbf{(D)} Levenshtein distance between the prediction and the masked word. \textbf{(I)} A small Levenshtein distance between the prediction and the masked word could indicate relevance. \\ 
  
  F9 &
  \textbf{(T)} Numeric \textbf{(D)} Semantic similarity computed as cosine similarity over word embeddings. \textbf{(I)} A prediction that is close in meaning to the masked word could have a higher likelihood of being relevant. \\
  
  F10$^{*}$ &
  \textbf{(T)} Ordinal \textbf{(D)} A value between zero and nine, indicating how frequently the prediction (in lemmatized form) appears across \emph{all BERT-generated predictions} over a given RS. \textbf{(I)} A smaller value could indicate a higher likelihood of relevance.\\ 
  
  F11$^{*}$\,$^{\dagger}$ &
  \textbf{(T)} Ordinal \textbf{(D)} A value between zero and nine, indicating how frequently the prediction (in lemmatized form) appears in the \emph{domain-specific corpus}. \textbf{(I)} A smaller value could indicate a higher likelihood of relevance.\\ 
  
  F12$^{\dagger}$\,$^{\ddagger}$ &
  \textbf{(T)} Numeric \textbf{(D)} Average TF-IDF rank of the prediction across all articles in the domain-specific corpus. \textbf{(I)} A higher rank could indicate a higher likelihood of relevance. \\ 
  
  F13$^{\dagger}$\,$^{\ddagger}$ &
  \textbf{(T)} Numeric \textbf{(D)} Maximum TF-IDF rank of the prediction across all articles in the domain-specific corpus. \textbf{(I)} Same intuition as that for F12.\\ 
  \hline
  \bottomrule
\end{tabularx}

\vspace*{.5em}
{\it\fontsize{8.5}{9}\selectfont
$^{*}$Zero is most frequent (top ten percentile) and nine is least frequent (bottom ten percentile).
$^{\dagger}$Feature uses domain-specific corpus.
$^{\ddagger}$TF-IDF values are normalized by Euclidean norm.}
\vspace*{-1.5em}
\end{table}

\sectopic{Step~6)~Filtering Noise from Predictions}
The predictions from Step 3 are noisy (i.e., have many false positives). To reduce the noise, we subject the predictions to a pre-trained ML-based filter. The most accurate ML algorithm for this purpose is selected empirically (see RQ2 in Section~\ref{sec:eval}). The selected algorithm is trained on the development and training portion of our dataset ($P_1$ in Table~\ref{tbl:dataset}, as we discuss in Section~\ref{sec:eval}). Due to our features in Table~\ref{tbl:featurestable} being generic and normalized, the resulting ML model can be used as-is over unseen documents without re-training (see RQ3 in Section~\ref{sec:eval} for evaluation of effectiveness). The output of this step is the list of BERT predictions that are classified as ``relevant'' by our filter; duplicates are excluded from the final results.

\section{Evaluation}\label{sec:eval}
In this section, we empirically evaluate our approach. During the process, we also build the pre-trained ML model required by Step~6 of the approach (Fig~\ref{fig:approach}).

\subsection{Research Questions (RQs)}
Our evaluation answers the following RQs using part of the PURE dataset~\cite{ferrari2017pure}. In lieu of expert input about incompleteness for the documents in this dataset, we apply the withholding strategy discussed in Section~\ref{sec:intro} to simulate incompleteness.

\sectopic{RQ1. How accurately can BERT predict relevant but missing terminology for an input RS?} The number of predictions that BERT generates per mask is a configurable parameter. RQ1 examines what value for this parameter offers the best trade-off for producing useful recommendations.

\sectopic{RQ2. Which ML classification algorithm most accurately filters unuseful predictions made by BERT?}
Useful recommendations from BERT come alongside a considerable amount of noise. In RQ2, we examine different ML algorithms to filter noise. We further study the impact of data balancing and cost-sensitive learning to prevent over-filtering. 

\sectopic{RQ3. How accurate are the recommendations generated by our approach over unseen documents?}
In RQ3, we combine the best BERT configuration from RQ1 with the filter models built in RQ2, and measure the accuracy of this combination over unseen data.

\subsection{Implementation and Availability}
We have implemented our approach in Python. 
The NLP pipeline is implemented using SpaCy 3.2.2.
For extracting word embeddings, we use GloVe~\cite{pennington2014glove}.
To obtain masked language model predictions from BERT, we use the Transformers 4.16.2 library by Hugging Face (\url{https://huggingface.co/}) and operated in PyTorch 1.10.2+cu113. Our ML-based filters are implemented in WEKA 3-8-5~\cite{frank2016weka}. To implement the ML features listed in Table~\ref{tbl:featurestable}, we use standard implementations of cosine similarity (over word embeddings) and Levenshtein distance~\cite{manning2008introduction}. The TFIDF-based features in this table (F12-13) use TfidfVectorizer from scikit-learn 1.0.2. Our implementation and evaluation artifacts are publicly available~\cite{Luitel:22}.

\subsection{Dataset}

Our evaluation is based on 40 documents from the PURE dataset~\cite{ferrari2017pure} -- a collection of public-domain requirements specifications. Many of the documents in PURE require manual cleanup (e.g., removal of table of contents, headers, section markers, etc.) We found 40 to be a good compromise between the effort that we needed to spend on cleanup and having a dataset large enough for statistical significance testing, mitigating the effects of random variation, and training ML-based filters. The selected documents, listed in Table~\ref{tbl:dataset}, cover 15 domains.
We partition the documents into two (disjoint) subsets $P_1$ and $P_2$. $P_1$ is used for approach development and tuning, i.e., for answering RQ1 and RQ2.
$P_2$, i.e., the documents unseen during development and tuning, is used for answering RQ3.
Our procedure for assigning documents to $P_1$ or $P_2$ was as follows: We first randomly selected one document per domain and put it into $P_2$; this is to maximize domain representation in RQ3. From the rest, we randomly selected 20 documents for inclusion in $P_1$, while attempting to have $P_1$ represent half of the data in terms of token count. Any remaining document after this process was assigned to $P_2$, thus giving us 20 documents in $P_2$ as well. Table~\ref{tbl:dataset} provides domain information and summary statistics for \hbox{documents in $P_1$ and $P_2$ after cleanup.}

\begin{table}[!t]
    \caption{Our Dataset (Subset of PURE~\cite{ferrari2017pure}). $P_1$ is for development and training and $P_2$ for testing.}
    \vspace*{-1em}
   \includegraphics[width=1\linewidth]{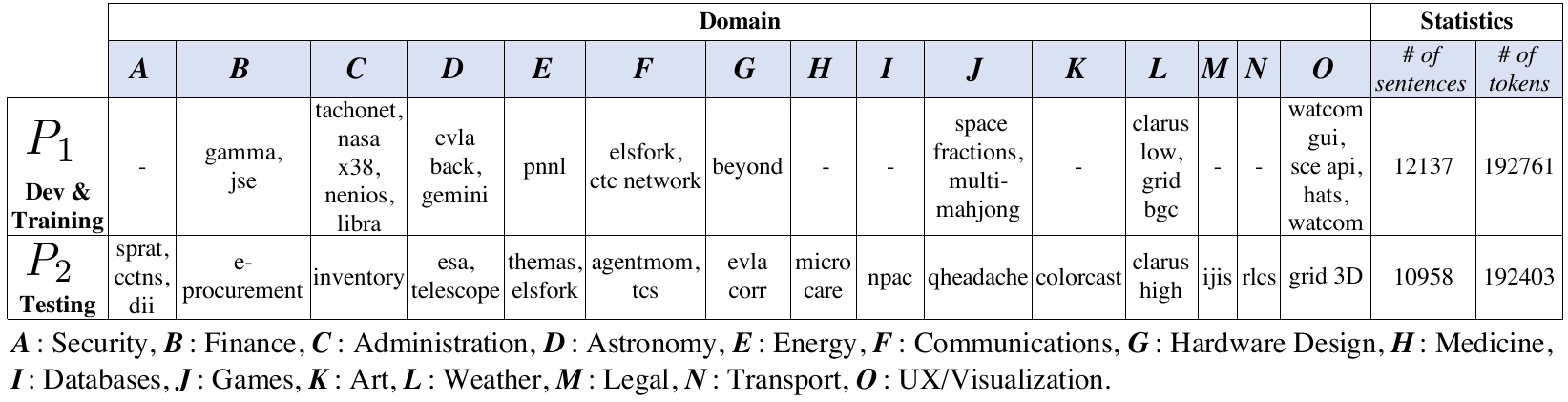}
    \label{tbl:dataset}
    \vspace*{-1em}
\end{table}

\subsection{Analysis Procedure}\label{subsec:proc}

\sectopic{EXPI.} This experiment answers RQ1. For every document $p\in P_1$, we randomly partition the set of sentences in $p$ into two subsets of (almost) equal sizes. In line with our arguments in Section~\ref{sec:intro}, we \emph{disclose} one of these subsets to BERT and \emph{withhold} the other. We apply Steps~1,~2~and~3 of our approach (Fig.~\ref{fig:approach}) to the disclosed half, \emph{as if this half were the entire input document}. We run Step~2 of our approach with four different numbers of predictions per mask: 5, 10, 15, and 20. For every document $p\in P_1$, we compute two metrics, \emph{Accuracy} and \emph{Coverage}, defined in Section~\ref{subsec:metricS}. As the number of predictions per mask increases from 5 to 20, the predictions made by BERT reveal more terms that are relevant to the withheld half. Nevertheless, as we will see in Section~\ref{subsec:results}, the benefits diminish beyond 15 predictions per mask.

To ensure that the trends we see as we increase the number of predictions per mask are not due to random variation, we pick different shuffles of each document $p$ across different numbers of predictions per mask. For example, the disclosed and withheld portions for a given document $p$ when experimenting with 5 predictions per mask are different random subsets than when experimenting with 10 predictions per mask.

\sectopic{EXPII.} This experiment answers RQ2 and further constructs the training set for the ML classifier in Step~6 of our approach (Fig.~\ref{fig:approach}). We recall the disclosed and withheld halves as defined in EXPI. For every document $p\in P_1$, we label predictions as ``relevant'' or ``non-relevant'' using the following procedure: Any prediction matching some term in the withheld half is labelled ``relevant''. The criterion for deciding whether two terms match is a cosine similarity of $\geq85$\% over GloVe word embeddings (introduced in Section~\ref{sec:background}). All other predictions are labelled ``non-relevant''. The conservative threshold of 85\%  ensures that only terms with the same lemma or with very high semantic similarity are matched. For each prediction, a set of features is calculated as detailed in Step~5 of our approach. It is paramount to note that Step~4, which is a prerequisite to Step~5, \emph{exclusively} uses the content of the disclosed half without any knowledge of the withheld half. The above process produces labelled data for each $p\in P_1$. We aggregate all the labelled data into a single \emph{training set}. This is possible because our features (listed in Table~\ref{tbl:featurestable}) are generic and normalized.

Equipped with a training set, we compare five widely used ML algorithms: Feed Forward Neural Network (NN), Decision Tree (DT), Logistic Regression (LR), Random Forest (RF)  and Support Vector Machine (SVM). All algorithms are tuned with optimal hyperparameters that maximize classification accuracy over the training set. For tuning, we apply multisearch hyperparameter optimization using random search~\cite{bergstra2012random}. The basis for tuning and comparing algorithms is ten-fold cross validation. We experiment with under-sampling the ``non-relevant'' class with and without cost-sensitive learning (CSL); the motivation is reducing false negatives (i.e., relevant terms incorrectly classified as ``non-relevant''). For CSL, we assign double the cost (penalty) to false negatives compared to false positives (i.e., noise). We further assess the importance of our features using information gain (IG)~\cite{WittenFrankEtAl17}. In our context, IG measures how efficient a given feature is in discriminating ``non-relevant'' from ``relevant'' predictions. A higher IG value implies a higher discriminative power.

\sectopic{EXPIII.} This experiment answers RQ3 by applying our end-to-end approach to unseen requirements documents, i.e., $P_2$. To conduct EXPIII, we need a pre-trained classifier for Step~6 of our approach (Fig.~\ref{fig:approach}). This classifier needs to be completely independent of $P_2$. We build this classifier using the training set derived from $P_1$, as discussed in EXPII. EXPIII follows the same strategy as in EXPI, which is to randomly withhold half of each document $p$ (now in $P_2$ rather than in $P_1$) and attempting to predict the novel terms of the withheld half. In contrast to EXPI, in EXPIII, predictions made by BERT are post-processed by a filter aimed at reducing noise. We repeat EXPIII \emph{five times} for each $p \in P_2$. This mitigates random variation resulting from the random selection of the disclosed and withheld halves, thus yielding more realistic ranges for performance. In EXPIII, we study three levels of filtering. Noting that there are 20 documents in $P_2$, the results reported for EXPIII use $20*5*3 = 300$ runs of our approach.

\subsection{Metrics}\label{subsec:metricS}
We define separate metrics for measuring (1) the quality of term predictions and (2)~the performance of filtering. The first set of metrics is used in RQ1 and RQ3 and the second set is used in RQ2 and RQ3.
To define our metrics, we need to introduce some notation. Let \hbox{$\mathsf{Lem}:\textsf{bag}\rightarrow \textsf{bag}$} be a function that takes a bag of words and returns another bag of words by \emph{lemmatizing} every element in the input bag. Let $\mathsf{U}: \textsf{bag}\rightarrow \textsf{set}$ be a function that removes duplicates from a bag and returns a set. Let $C$ denote the set of common words and stopwords as explained under Step~3 in Section~\ref{sec:approach}. Given a document $p$ treated as a bag of words, the terminological content of $p$'s disclosed half, denoted $h_1$, is given by set  $X=\mathsf{U}(\mathsf{Lem}(h_1))$. In a similar vein, the terminological content of $p$'s withheld half, denoted $h_2$, is given by \hbox{set $Y=\mathsf{U}(\mathsf{Lem}(h_2))$}. What we would like to achieve through BERT is to predict as much of the \emph{novel} terminology in the withheld half as possible. This novel terminology can be defined as set \hbox{$N = (Y - X) - C$}. Let bag $V$ be the output of Step~3  (Fig.~\ref{fig:approach}) when the approach is applied \emph{exclusively} to the disclosed half of a given document (i.e., $h_1$). Note that $V$ is already free of any terminology that appears in the disclosed half, as well as of \hbox{all common words and stopwords.}

\sectopic{Quality of term predictions.}
Let set $D$ denote the (duplicate-free) lemmatized predictions that have the potential to hint at novel terminology in the withheld half of a given document. Formally, let $D= \textsf{U}(\textsf{Lem}(V))$. We define two metrics, \emph{Accuracy} and \emph{Coverage} to measure the quality of $D$.
\emph{Accuracy} is the ratio of terms in $D$ matching some term in $N$, to the total number of terms in $D$. That is, $\mathit{Accuracy} = |\{t \in D \mid t\ \text{matches some}\ t' \in N\}|/|D|$. A term $t$ matches another term $t'$ if the word embeddings have a cosine similarity of $\geq 85\%$ (already discussed under EXPII in Section~\ref{subsec:proc}).
The second metric, \emph{Coverage}, is defined as the ratio of terms in $N$ matching some term in $D$, to the total number of terms in $N$. That is, $\mathit{Coverage} = |\{t \in N \mid t\ \text{matches some}\ t' \in D\}|/|N|$. The intuition for Accuracy and Coverage is the same as that for the standard Precision and Recall metrics, respectively. Nevertheless, since our matching is inexact and based on a similarity threshold, it is possible for more than one term in $D$ to match an individual term in $N$. Coverage, as we define it, excludes multiple matches, providing a measure of how much of the novel terminology in the withheld half is hinted at by BERT.

\sectopic{Quality of filtering.}
As explained earlier, our filter is a binary classifier to distinguish relevance and non-relevance for the outputs from BERT. To measure filtering performance, we use the standard metrics of \emph{Classification Accuracy}, \emph{Precision} and \emph{Recall}. True positive (TP), false positive (FP), true negative (TN) and false negative (FN) are defined as follows: A TP is a classification of ``relevant'' for a term that has a match in set $N$ (defined earlier). A FP is a classification of ``relevant'' for a term that does \emph{not} have a match in $N$. A TN is a classification of ``non-relevant'' for a term that does not have a match in $N$. A FN is a classification of ``non-relevant'' for a term that does have a match in $N$. \emph{Classification Accuracy} is calculated as $(TP+TN)/(TP+TN+FP+FN)$. \emph {Precision} is calculated as $TP/(TP+FP)$ and \emph{Recall} as $TP/(TP+FN)$.

\subsection{Results}\label{subsec:results}
\begin{figure}
\centering
\includegraphics[width=1\linewidth]{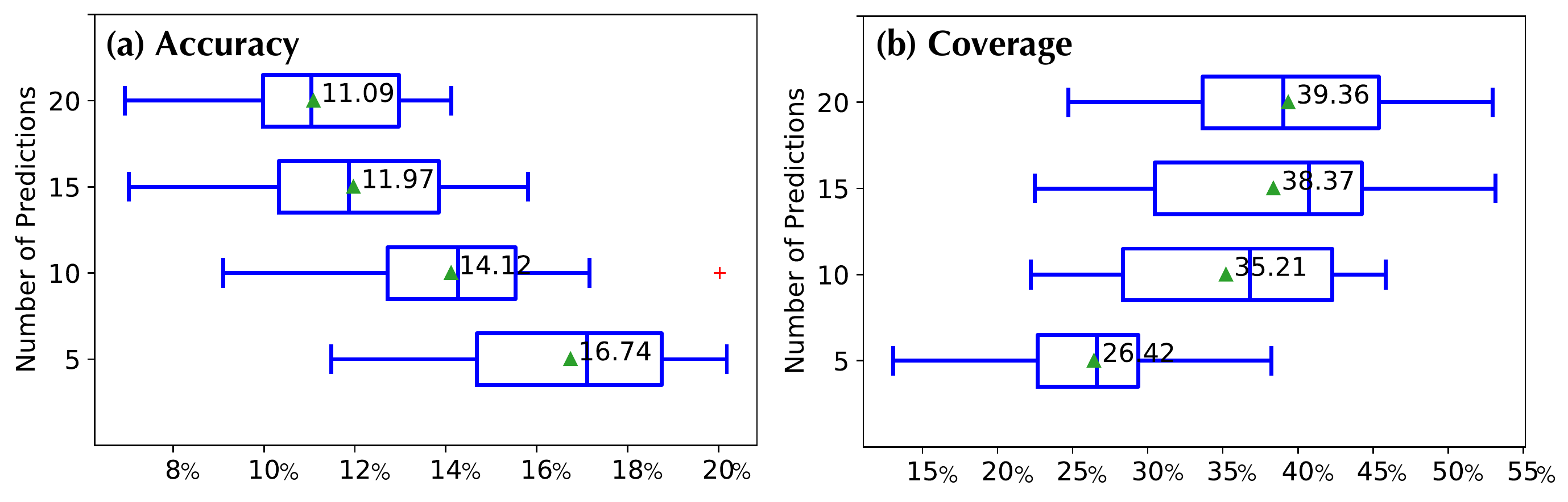}
\vspace*{-2.5em}
\caption{(a)~Accuracy  and (b)~Coverage  for Different Numbers of Predictions per Mask. Each box plot represents 20 datapoints (one datapoint per $p\in P_1$) as computed by EXPI in Section~\ref{subsec:proc}.}\label{fig:RQ1-Accuracy-Coverage}
\end{figure}

\sectopic{RQ1.} Fig.~\ref{fig:RQ1-Accuracy-Coverage} provides box plots for Accuracy and Coverage with the number of predictions by BERT ranging from 5 to 20 in increments of 5. Each box plot is based on 20 datapoints; each datapoint represents one document in $P_1$. We perform statistical significance tests on the obtained metrics using the non-parametric pairwise Wilcoxon's rank sum test~\cite{capon1991elementary} and Vargha-Delaney's effect size~\cite{vargha2000critique}. Table~\ref{fig:Statistical Significance Testing} shows the results of the statistical tests. Each column in the table compares Accuracy and Coverage across two levels of predictions per mask. For example, the \emph{5 vs. 10} column compares the metrics for when BERT generates 5 predictions per mask versus when it generates 10.

For Accuracy, Fig.~\ref{fig:RQ1-Accuracy-Coverage} shows a downward trend as the number of predictions per mask increases. Based on Table~\ref{fig:Statistical Significance Testing}, the decline in Accuracy is statistically significant with each increase in the number of predictions, the exception being the increase from 15 to 20, where the decline is not statistically significant. 

For Coverage, Fig.~\ref{fig:RQ1-Accuracy-Coverage} shows an upward but saturating trend. Five predictions per mask is too few: all other levels are significantly better. Twenty is too many, notably because of the lack of a significant difference for Coverage in the \hbox{\emph{10 vs. 20}} column of Table~\ref{fig:Statistical Significance Testing}. The choice is thus between 10 and 15. We select 15 as this yields an average increase of 3.2\% in Coverage compared to 10 predictions per mask. This increase is not statistically significant. Nevertheless, the price to pay is an average decrease of $(14.12 - 11.97 =)\ 2.15\%$ in Accuracy. Given the importance of Coverage, we deem 15 to be a better compromise than 10.

\begin{table}
\caption{Statistical Significance Testing for the Results of Fig.~\ref{fig:RQ1-Accuracy-Coverage}.}
\label{fig:Statistical Significance Testing}
\vspace*{-.5em}
\centering
\includegraphics[width=1\linewidth]{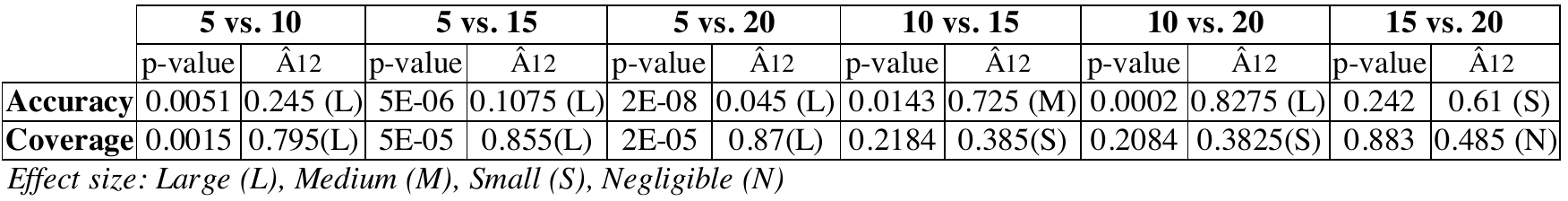}
\end{table}

\vspace*{.5em}
\begin{samepage}
\begin{mdframed}[style=MyFrame]
\emph{The answer to {\bf RQ1} is: When requirements omissions are simulated by withholding, having BERT make 15 predictions per mask is the best trade-off for detecting missing terminology. BERT predicts terms that, on average, hint at $\approx$4 out of 10 omissions (Coverage $\approx$38\%). On average, $\approx$1 in 8 predictions is relevant (Accuracy $\approx$12\%).}
\end{mdframed}
\end{samepage}
\vspace*{.5em}

\begin{table}[!b]
    \caption{ML Algorithm Selection (RQ2). All algorithms have tuned hyperparameters.}
    \label{fig:SecondRQ}
    \vspace*{-.75em}
   \includegraphics[width=1\linewidth]{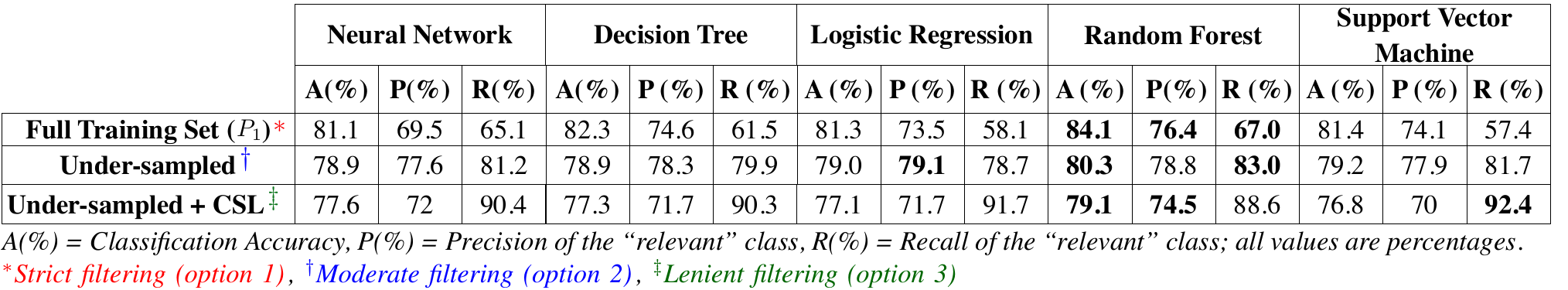}
   \vspace*{-2em}
\end{table}

\begin{wrapfigure}{r}{0.44\columnwidth}
  \centering
  \includegraphics[width=0.44\columnwidth]{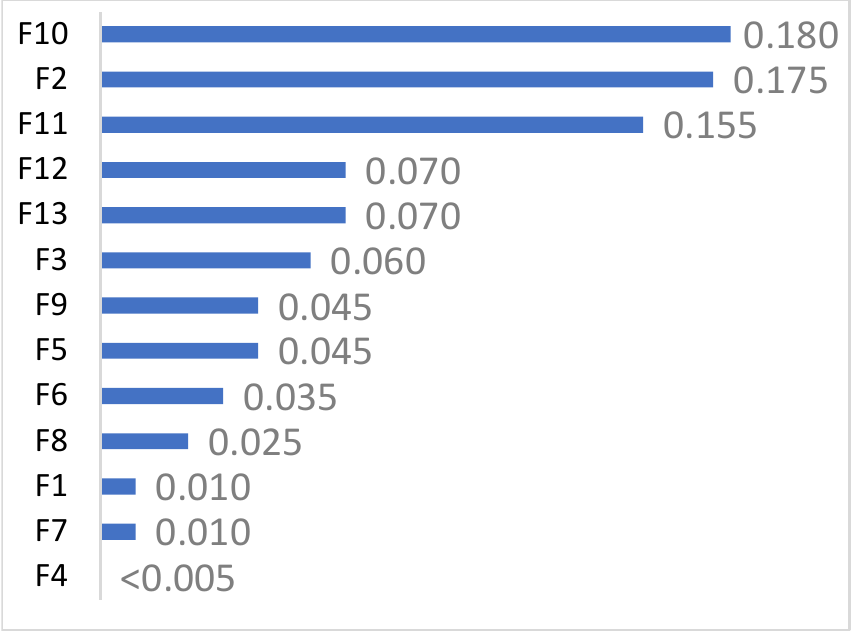}
  \vspace{-2em}
  \caption{Feature Importance (Avg).}
\label{fig:Information Gain}
  \vspace{-1.5em}
\end{wrapfigure}

\sectopic{RQ2.} Table~\ref{fig:SecondRQ} shows the results for ML-algorithm selection using the full ($P_1$) training set (61,996 datapoints), the under-sampled training set (36,842 datapoints), and the under-sampled training set alongside CSL. Classification Accuracy, Precision and Recall are calculated using ten-fold cross validation. In the table, we highlight the best result for each metric in bold. When one uses the full training set  (\emph{option~1}) or the under-sampled training set without CSL (\emph{option~2}), Random Forest (RF) turns out to be the best alternative. When the under-sampled training set is combined with CSL (\emph{option~3}), RF still has the best Accuracy and Precision. However, Support Vector Machine (SVM) presents a moderate advantage in terms of Recall. Since option~3 is meant at further improving the filter's Recall, we pick SVM as the best alternative for this particular option. Fig.~\ref{fig:Information Gain} lists the features of Table~\ref{tbl:featurestable} in descending order of 
information gain (IG), averaged across options 1, 2 and 3. We observe that our corpus-based features (F11--F13) are among the most influential, thus justifying the use of a domain-specific corpus extractor in our approach.

Compared to option~1, options~2~and~3 get progressively more lax by filtering \emph{less}. We answer RQ3 using RF for options 1 and 2 and SVM for option 3. For better intuition, we refer to option 1 as \emph{strict}, option 2 as \emph{moderate} and option 3 as \emph{lenient}.

\vspace*{.5em}
\begin{samepage}
\begin{mdframed}[style=MyFrame]
\emph{The answer to {\bf RQ2} is: RF and SVM yield the most accurate filter for unuseful predictions. RF is a better alternative for more aggressive filtering, whereas SVM is a better alternative for more lax filtering \hbox{(thus better preserving Recall).}}
\end{mdframed}
\end{samepage}
\vspace*{.5em}

\begin{wrapfigure}{r}{0.5\columnwidth}
  \vspace*{-1.5em}
  \centering
\includegraphics[width=0.5\columnwidth]{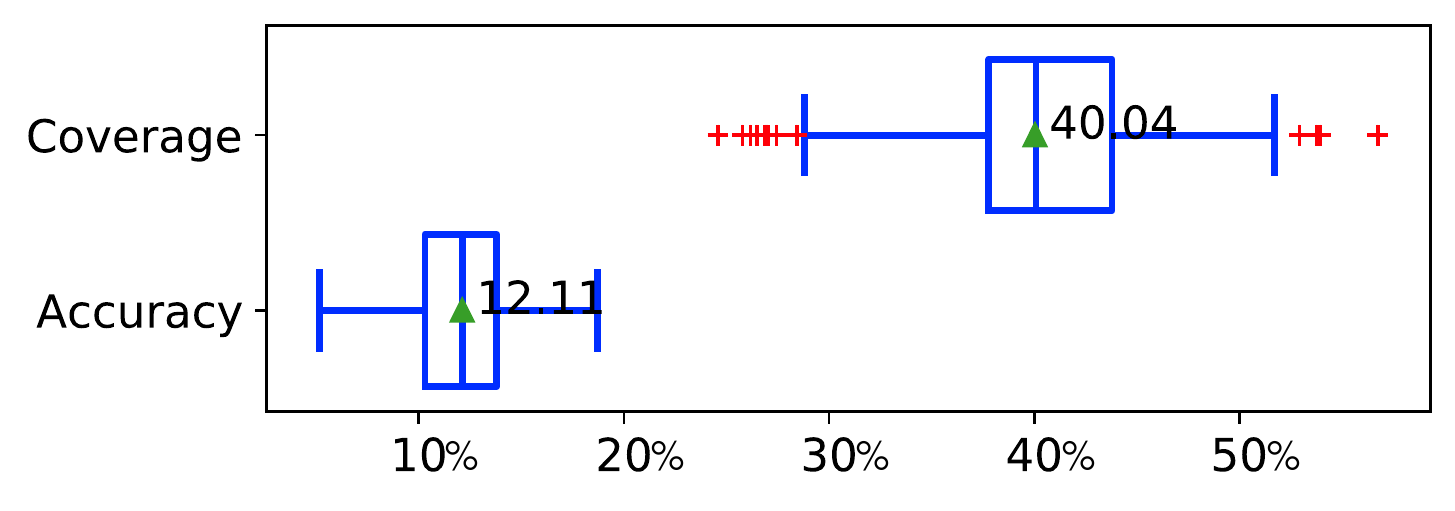}
  \caption{Accuracy and Coverage over Test Set ($P_2)$ without Filtering.}
  \vspace{-1.5em}
 \label{fig:RQ3-NoFilter}
\end{wrapfigure}
\sectopic{RQ3.} Without filters and over our test set ($P_2$ in Table~\ref{tbl:dataset}), the 15 predictions per mask made by BERT have an average Accuracy of 12.11\%  and average Coverage of 40.04\%. Box plots are provided in Fig.~\ref{fig:RQ3-NoFilter}. We recall from Section~\ref{subsec:proc} (EXPIII) that five different random shuffles are performed for each $p\in P_2$. The plots in Fig.~\ref{fig:RQ3-NoFilter} are based on $5*20=100$ runs.

\begin{figure}[!t]
\centering
    \hspace*{-2em}\mbox{}\includegraphics[width=1.05\linewidth]{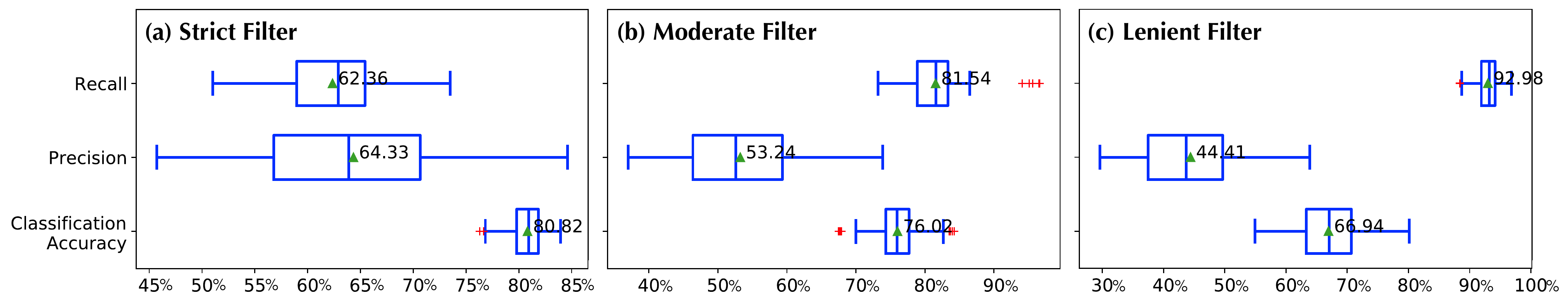}
\vspace*{-2.5em}
\caption{Filtering Classification Accuracy, Precision and Recall over Test Set ($P_2$).}
\vspace*{2em}
\label{fig:RQ3-PartA}
\end{figure}

Fig.~\ref{fig:RQ3-PartA} shows the performance of 
our three filters, namely strict, moderate, and lenient, over $P_2$. We observe that for all three filters, Precision levels over unseen data are lower than the cross-validation results in RQ2 (Table~\ref{fig:SecondRQ}). This discrepancy was particularly expected for the moderate and lenient filters, noting that, for these two filters, Table~\ref{fig:SecondRQ} reports performance over an under-sampled dataset. As for Recall, the results of Fig.~\ref{fig:RQ3-PartA} indicate that the filters behave consistently with trends seen in cross-validation. This consistency provides evidence that the filters did not overfit to the training data and are thus sufficiently generalizable.

\begin{figure}[!t]
\centering\mbox{\hspace*{-2em}}
    \includegraphics[width=1.03\linewidth]{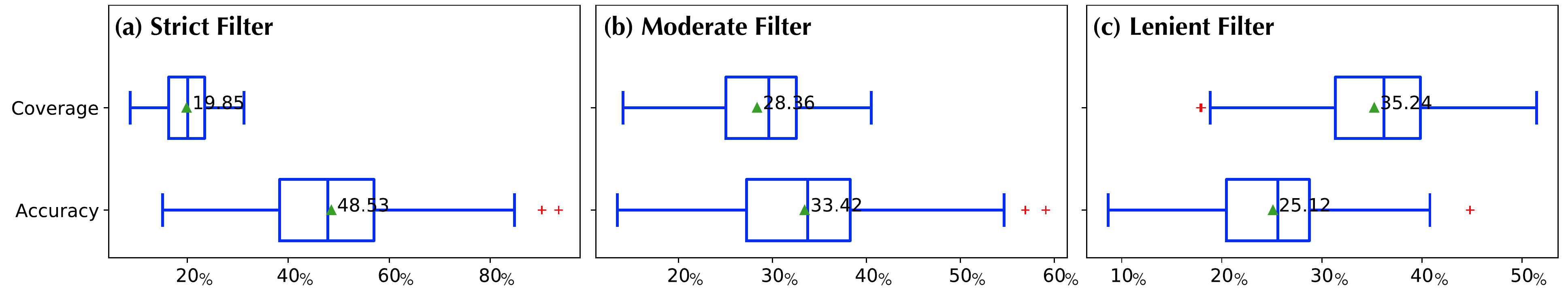}
\vspace*{-1.5em}
\caption{Accuracy and Coverage over Test Set with (a)\,Strict, (b)\,Moderate, and (c)\,Lenient Filter.}
\label{fig:RQ3-PartB}
\end{figure}

Fig.~\ref{fig:RQ3-PartB} shows the word-prediction Accuracy and Coverage results \emph{after} filtering. Which filtering option the user selects depends on how the user wishes to balance the overhead of reviewing non-relevant recommendations against potentially finding a larger number of relevant terms missing from requirements.

The lenient filter increases Accuracy by an average $\approx$13\% while decreasing Coverage by $\approx$5\%. The strict filter increases Accuracy by an average $\approx$36\% while decreasing Coverage by $\approx$20\%. All filters impact both Accuracy and Coverage in a statistically significantly manner with medium to large effect sizes.

\begin{samepage}
\begin{mdframed}[style=MyFrame]
\emph{The answer to {\bf RQ3} is: Depending on how aggressively one chooses to filter noise from BERT's masked-word predictions, the average Accuracy of our approach ranges between $\approx$49\% and $\approx$25\%. With a strict filter, approximately one in two recommendations made by our approach is relevant, whereas with a lenient filter, approximately one in four is. With a lenient filter, the recommendations hint at $\approx$35\% of the (simulated) missing terminology. With a strict filter, this number decreases to $\approx$20\%.}
\end{mdframed}
\end{samepage}

\subsection{Limitations and Validity Considerations}\label{subsec:threats}

\sectopic{Limitations.} We did not have access to domain experts for identifying genuine cases of incompleteness.
Our evaluation therefore simulates incompleteness by withholding content from existing requirements. Future user studies remain necessary for developing a qualitative interpretation of our results and drawing more definitive conclusions about the usefulness of our approach.

\sectopic{Validity Considerations.} The validity considerations most pertinent to our evaluation are internal, construct and conclusion validity. With regard to \emph{internal validity}, we note that the disclosed and withheld portions were chosen randomly. To mitigate random variation, we used a sizable dataset (40 documents) and further employed repeated experimentation (RQ3). With regard to \emph{construct validity}, we note that our metrics for term predictions discard any terms already seen in the disclosed portion as well as any duplicates, common words and stopwords. This helps ensure that our metrics provide a legitimate assessment of the quality of the predictions made by BERT. That being said, our current quality assessment is not based on human judgement, and instead hinges on an automatically calculated similarity measure. A human-based validation of the predictions by BERT is therefore essential for better gauging the practical value of our approach. With regard to \emph{conclusion validity}, we note that, we chose a 50-50 split between the disclosed and withheld portions. Assuming that the useful predictions by BERT are evenly distributed across the withheld portion, similar benefits should be seen with different split ratios, as long as the withheld portion is not excessively small. We anticipate that there will be a limit to how small the withheld portion can be, before it becomes too difficult to make relevant predictions. This limit determines the sensitivity of our approach to incompleteness. Further \hbox{research is required to establish this limit.}

\section{Related Work}\label{sec:related}

Ferrari et al.~\cite{ferrari2014measuring}
propose an NLP-based approach for automatically extracting relevant terms and relations from descriptions such as client-meeting transcripts. Based on the extracted results, they make recommendations for improving the completeness of requirements. 
Dalpiaz et al.~\cite{dalpiaz2018pinpointing} develop a technique based on NLP and visualization to explore commonalities and differences between multiple viewpoints and thereby help stakeholders pinpoint occurrences of ambiguity and incompleteness. In the above works, the sources of knowledge used for completeness checking (descriptions or alternative viewpoints) are existing development artifacts. Our work uses a generative language model, BERT, for completeness checking. In contrast to the above works, our approach does not assume the existence of any user-provided artifacts against which to compare the completeness of requirements.

Arora et al.~\cite{arora2019empirical} use domain models for detecting incompleteness in  requirements. The authors simulate requirements omissions and demonstrate that domain models can signal the presence of these omissions. Again, there is an assumption about the existence of an additional development artifact -- in this case, a domain model. This limits the applicability of the approach to when a sufficiently detailed domain model exists. 

Bhatia et al.~\cite{bhatia2018semantic} address incompleteness in privacy policies by representing data actions as semantic frames. They identify the expected semantic roles for a given frame, and consequently determine incompleteness by identifying missing role values. Cejas et al.~\cite{cejas2021ai} use NLP and ML for completeness checking of privacy policies. Their approach identifies instances of pre-defined concepts such as ``controller'' and ``legal basis'' in a given policy. It then verifies through rules whether all applicable concepts are covered. The above works deal with privacy policies only and have a predefined conceptual model for textual content. Our BERT-based approach is not restricted to a particular application domain and does not have a fixed conceptualization of the textual content under analysis. Instead, we utilize BERT's pre-training and attention mechanism to make contextualized recommendations for improving completeness.

Shen and Breaux~\cite{shendomain} propose an NLP-based approach for extracting domain knowledge from user-authored scenarios and word embeddings. While this approach is not concerned with checking the completeness of requirements, it uses BERT's MLM for generating alternatives by masking words in requirements statements. Our approach uses BERT's MLM in a similar manner. In contrast to this earlier work, we take steps to address the challenge arising from such use of BERT over requirements, namely the large number of non-relevant alternatives (false positives) generated. We propose a ML-based filter that uses a combination of NLP and statistics extracted from a domain-specific corpus to reduce the incidence of false positives.
\section{Conclusion}\label{sec:conclusion}
Our results indicate that masked-word predictions by BERT, when complemented with a mechanism to filter noise, have potential for detecting incompleteness in requirements. In future work, we plan to conduct user studies to more conclusively assess this potential. Other directions for future work include
experimentation with BERT variants and improving accuracy via fine-tuning.

\vspace*{.5em}\sectopic{Acknowledgements.} This work was funded by the Natural Sciences and Engineering Research Council of Canada (NSERC) under the Discovery and Discovery Accelerator programs. We are grateful to Shiva Nejati, Sallam Abualhaija and Jia Li for helpful discussions. We thank the anonymous reviewers of REFSQ~2023 for their constructive comments.
\bibliographystyle{unsrt}
\bibliography{references}

\begin{thebibliography}{10}

\bibitem{ferrari2017pure}
Alessio Ferrari, Giorgio~Oronzo Spagnolo, and Stefania Gnesi.
\newblock {PURE}: A dataset of public requirements documents.
\newblock In {\em RE}, 2017.

\bibitem{Zowghi:03}
Didar Zowghi and Vincenzo Gervasi.
\newblock The three {Cs} of requirements: Consistency, completeness, and
  correctness.
\newblock In {\em REFSQ}, 2003.

\bibitem{Zowghi:03b}
Didar Zowghi and Vincenzo Gervasi.
\newblock On the interplay between consistency, completeness, and correctness
  in requirements evolution.
\newblock {\em IST}, 45(14), 2003.

\bibitem{arora2019empirical}
Chetan Arora, Mehrdad Sabetzadeh, and Lionel Briand.
\newblock An empirical study on the potential usefulness of domain models for
  completeness checking of requirements.
\newblock {\em EMSE}, 24(4), 2019.

\bibitem{ferrari2014measuring}
Alessio Ferrari, Felice dell'Orletta, Giorgio~Oronzo Spagnolo, and Stefania
  Gnesi.
\newblock Measuring and improving the completeness of natural language
  requirements.
\newblock In Camille Salinesi and Inge van~de Weerd, editors, {\em REFSQ},
  2014.

\bibitem{dalpiaz2018pinpointing}
Fabiano Dalpiaz, Ivor van~der Schalk, and Garm Lucassen.
\newblock Pinpointing ambiguity and incompleteness in requirements engineering
  via information visualization and {NLP}.
\newblock In {\em Requirements Engineering: Foundation for Software Quality},
  2018.

\bibitem{devlin2018bert}
Jacob Devlin, Ming-Wei Chang, Kenton Lee, and Kristina Toutanova.
\newblock {BERT}: Pre-training of deep bidirectional transformers for language
  understanding.
\newblock In {\em NAACL-HLT}, 2019.

\bibitem{Luitel:22}
Dipeeka Luitel, Shabnam Hassani, and Mehrdad Sabetzadeh.
\newblock Replication package, 2023.
\newblock \url{https://doi.org/10.6084/m9.figshare.22041341}.

\bibitem{jurafsky2009speech}
Daniel Jurafsky and J.~Martin.
\newblock {\em Speech and Language Processing}.
\newblock Prentice Hall, 2 edition, 2009.

\bibitem{hey2020norbert}
Tobias Hey, Jan Keim, Anne Koziolek, and Walter~F. Tichy.
\newblock {NoRBERT}: Transfer learning for requirements classification.
\newblock In {\em RE}, 2020.

\bibitem{ezzini2022automated}
Saad Ezzini, Sallam Abualhaija, Chetan Arora, and Mehrdad Sabetzadeh.
\newblock Automated handling of anaphoric ambiguity in requirements: A
  multi-solution study.
\newblock In {\em ICSE}, 2022.

\bibitem{mikolov2013linguistic}
Tomas Mikolov, Wen-tau Yih, and Geoffrey Zweig.
\newblock Linguistic regularities in continuous space word representations.
\newblock In {\em NAACL-HLT}, 2013.

\bibitem{pennington2014glove}
Jeffrey Pennington, Richard Socher, and Christopher Manning.
\newblock {G}lo{V}e: Global vectors for word representation.
\newblock In {\em EMNLP}, 2014.

\bibitem{WittenFrankEtAl17}
Ian~H. Witten, Eibe Frank, and Mark~A. Hall.
\newblock {\em Data Mining: Practical Machine Learning Tools and Techniques}.
\newblock Morgan Kaufmann, 4 edition, 2017.

\bibitem{berry2017}
Daniel~M. Berry, Jane Cleland-Huang, Alessio Ferrari, Walid Maalej, John
  Mylopoulos, and Didar Zowghi.
\newblock Panel: context-dependent evaluation of tools for {NL RE} tasks:
  recall vs. precision, and beyond.
\newblock In {\em RE}, 2017.

\bibitem{ezzini2021using}
Saad Ezzini, Sallam Abualhaija, Chetan Arora, Mehrdad Sabetzadeh, and Lionel
  Briand.
\newblock Using domain-specific corpora for improved handling of ambiguity in
  requirements.
\newblock In {\em ICSE}, 2021.

\bibitem{cui2008corpus}
Gaoying Cui, Qin Lu, Wenjie Li, and Yi-Rong Chen.
\newblock Corpus exploitation from {Wikipedia} for ontology construction.
\newblock In {\em LREC}, 2008.

\bibitem{ferrari2017detecting}
Alessio Ferrari, Beatrice Donati, and Stefania Gnesi.
\newblock Detecting domain-specific ambiguities: an {NLP} approach based on
  {Wikipedia} crawling and word embeddings.
\newblock In {\em AIRE}, 2017.

\bibitem{ezzini2022wikidominer}
Saad Ezzini, Sallam Abualhaija, and Mehrdad Sabetzadeh.
\newblock {WikiDoMiner}: {Wikipedia} domain-specific miner.
\newblock In {\em ESEC/FSE}, 2022.

\bibitem{Berry:03}
Daniel~M. Berry, Erik Kamsties, and Michael Krieger.
\newblock From contract drafting to software specification: Linguistic sources
  of ambiguity, a handbook, 2003.

\bibitem{arora2015tse}
Chetan Arora, Mehrdad Sabetzadeh, Lionel Briand, and Frank Zimmer.
\newblock Automated checking of conformance to requirements templates using
  natural language processing.
\newblock {\em IEEE TSE}, 41(10), 2015.

\bibitem{arora2017tse}
Chetan Arora, Mehrdad Sabetzadeh, Lionel Briand, and Frank Zimmer.
\newblock Automated extraction and clustering of requirements glossary terms.
\newblock {\em IEEE TSE}, 43(10), 2017.

\bibitem{frank2016weka}
Ian~H. Witten, Eibe Frank, Mark~A. Hall, and Christopher~J. Pal.
\newblock {\em The {WEKA} Workbench. Online Appendix for "Data Mining:
  Practical Machine Learning Tools and Techniques"}.
\newblock Morgan Kaufmann Publishers Inc., 4th edition, 2016.

\bibitem{manning2008introduction}
Christopher Manning, Prabhakar Raghavan, and Hinrich Schütze.
\newblock {\em Introduction to Information Retrieval}.
\newblock Syngress, 2008.

\bibitem{bergstra2012random}
James Bergstra and Yoshua Bengio.
\newblock Random search for hyper-parameter optimization.
\newblock {\em JMLR}, 13(2), 2012.

\bibitem{capon1991elementary}
J.~Anthony Capon.
\newblock {\em Elementary Statistics for the Social Sciences: Study Guide}.
\newblock Wadsworth, 1991.

\bibitem{vargha2000critique}
Andras Vargha and Harold Delaney.
\newblock A critique and improvement of the {CL} common language effect size
  statistics of {McGraw and Wong}.
\newblock {\em Journal of Educational and Behavioral Statistics}, 25(2), 2000.

\bibitem{bhatia2018semantic}
Jaspreet Bhatia and Travis Breaux.
\newblock Semantic incompleteness in privacy policy goals.
\newblock In {\em RE}, 2018.

\bibitem{cejas2021ai}
Orlando Amaral~Cejas, Sallam Abualhaija, Damiano Torre, Mehrdad Sabetzadeh, and
  Lionel Briand.
\newblock {AI}-enabled automation for completeness checking of privacy
  policies.
\newblock {\em IEEE TSE}, 48(11), 2022.

\bibitem{shendomain}
Yuchen Shen and Travis Breaux.
\newblock Domain model extraction from user-authored scenarios and word
  embeddings.
\newblock In {\em AIRE}, 2022.

\end{thebibliography}
\end{document}